\documentclass[final]{aipproc}
\usepackage{amsmath}
\layoutstyle{8x11single}

\begin{document}
\title{Towards a New Minimal $SO(10)$ Unification}
\author{Stefano Bertolini}{address={INFN, Sezione di Trieste, Italy}}
\author{Luca Di Luzio}{address={Institut f\"{u}r Theoretische Teilchenphysik,
Karlsruhe Institute of Technology (KIT), D-76128 Karlsruhe, Germany}}
\author{Michal Malinsk\'y\footnote{Presenting author.}\,\,\,}{address={AHEP Group, Instituto de F\'{\i}sica Corpuscular -- C.S.I.C./Universitat de Val\`encia, Edificio de Institutos de Paterna, Apartado 22085, E 46071 Val\`encia, Spain}}
\begin{abstract}
We argue that non-supersymmetric $SO(10)$ models based on a renormalizable Higgs sector in which spontaneous symmetry breaking is triggered by the VEVs of a $45$-dimensional adjoint and a $126$-dimensional tensor representations can provide a potentially realistic yet relatively simple framework for a future robust estimate of the proton lifetime. Following closely the work~\cite{Bertolini:2012im} we comment on the gauge unification constraints on the $B-L$  breaking scale and show that there are several regions in the parameter space of the minimal model where the seesaw scale in the phenomenologically favoured ballpark of around $10^{13\div 14}$~GeV is consistently supported.
\end{abstract} 
\maketitle

\section{Introduction}
Though the Grand Unification paradigm is around for almost four decades \cite{Georgi:1974sy,Fritzsch:1974nn} it still attracts a lot of attention across the high-energy-physics community. Unlike many other extensions of the Standard Model (SM) featuring extra symmetries (global or local), Grand Unified Theories (GUTs) tell us implicitly something about the typical scale the extended gauge symmetry such as $SU(5)$ or $SO(10)$ is spontaneously broken and, at the same time, the underlying gauge dynamics always leaves its imprints at low energies. The most prominent of these are namely the heavy topologically stable monopoles created at the GUT phase transition in the early Universe and the baryon and lepton number non-conservation associated to the gauge and Yukawa interactions among quarks and leptons sharing irreducible representations of the GUT gauge group. 

Although it may be fundamentally difficult to get a grip on the GUT monopoles that could have been extremely diluted during the inflationary period, baryon and lepton number violating processes such as the proton decay, neutron-antineutron oscillations, neutrinoless double beta decay etc. can be within reach of upcoming facilities.    
In particular, proton instability - if observed - would provide a smoking-gun signal of Grand Unification\footnote{Barring few exceptions like e.g. R-parity violating supersymmetry etc.} and as such it has been searched for for decades by many dedicated experiments. 
Among these, the best sensitivity is provided by large-volume water \v{C}erenkov detectors (such as, e.g., IMB, Kamiokande, Super-Kamiokande) where the proton decay searches naturally support neutrino physics activities. It is also likely that in the near future these ``classical'' experiments will be further complemented by large liquid argon detectors (e.g. GLACIER, LArTPC) whose tracking properties and, in particular, a very good particle discrimination potential, can improve substantially the existing proton lifetime limits, especially for the decay modes with Kaons in the final state.   
Reaching a mega-ton scale (as assumed in the recent proposals such as Hyper-Kamiokande) is, however, a costly enterprise; a robust proton lifetime prediction including, optimally, branching fractions into at least the leading modes at a factor-of-few level of precision is vital to make a real physics case. 

From the theory point of view, reaching such a high level of accuracy within a specific GUT is often an enormous challenge. This has to do namely with the fact that, unlike the gauge structure dictated solely by the shape of the relevant gauge group the form of the Higgs sector is not derived from first principles and as such it is mainly subject to experimental (and often also aesthetic)  constraints. Since, at the same time, it supplies the key ingredients entering the proton lifetime calculation, namely, the symmetry breaking pattern constraining the position of the GUT scale and the Yukawa interactions governing the flavour structure of the relevant Feynman diagrams, it becomes one of the central issues of the whole GUT programme.   
In that respect, the quality of any such prediction is strongly correlated with the ``simplicity'' of a given GUT model; on the other hand, oversimplified settings are often trivially incompatible with the existing electroweak and flavour data. 
Remarkably enough, the borderline between ``trivially wrong'' and ``too complicated to be predictive''  models is rather thin and there are only several settings which historically qualified as ``truly minimal'' GUTs. 

\paragraph{Minimal $SU(5)$ models}  
Concerning the basic $SU(5)$ GUTs, neither the original Georgi-Glashow scenario \cite{Georgi:1974sy} nor its minimally supersymmetrized version provides such a minimal setting; while the former does not give the right weak mixing angle the latter generally predicts an overly fast $d=5$ proton decay~\cite{Murayama:2001ur}\footnote{For a recent attempt to save the minimal SUSY $SU(5)$ see, e.g., \cite{Martens:2010nm}.}. Moreover, both fail on the flavour side as they do not accommodate massive neutrinos. There are several ways out advocated in the existing literature: Giving up renormalizability the proton lifetime in the minimal supersymmetric $SU(5)$ can be, at least in principle, prolonged to acceptable values \cite{Bajc:2002pg,Bajc:2002bv}. On a similar footing, one can, e.g., add extra Higgs representations such as $15_{H}$ or $45_{H}$ and/or extra matter fields such as $24_{M}$ in attempt to make the flavour structure of the model more realistic (giving masses to neutrinos and, at the same time, smearing the notorious down-quark and charged-lepton mass degeneracy). At the same time, accidentally light multiplets can restore the gauge coupling unification even in the non-supersymmetric scenario. For more detail, an interested reader is deferred namely to the original works \cite{Dorsner:2005fq,Bajc:2006ia} and references therein. Unfortunately, these models in practice still rely on a certain (though limited) set of effective operators so their predictive power is not very clear. Thus, in what follows, we shall  stick to renormalizable GUTs.
\paragraph{Minimal $SO(10)$ models}
From the group theory point of view, the minimal Higgs sector\footnote{For an alternative approach exploiting a single 144-dimensional representation for the entire $SO(10)\to {\rm SM}$ symmetry breaking see e.g.\cite{Babu:2005gx}.} that can, at least in principle, lead to a full breakdown of the GUT to the SM gauge symmetry contains the rank-preserving adjoint representation $45_{H}$ supplemented by a rank-reducing complex representation, typically, $16_{H}$ or $126_{H}$. 
Unfortunately, this simple setting does not seem to work well in practice. 
As shown in \cite{Buccella:1980qb,Yasue:1980fy,Anastaze:1983zk,Babu:1984mz} the vacuum stability of the $45_{H}+16_{H}/126_{H}$ Higgs models requires a near $SU(5)$-like arrangement of the $45_{H}$ VEVs, in obvious conflict with the generic shape of the basic $SO(10)$ gauge unification patterns \cite{Chang:1984qr,Deshpande:1992em,Bertolini:2009qj}.
 
Similarly, in supersymmetry,  extra constraints from $F$- and $D$-flatness  align\footnote{For an interesting exception see, e.g., the minimal flipped $SO(10)$ scenario advocated in~\cite{Bertolini:2010yz}.} the pair of VEVs in $45_{H}$ along the $SU(5)$-preserving direction of the $16_{H}/126_{H}$ VEVs~\cite{Aulakh:2000sn}. Hence, the minimal renormalizable SUSY SO(10) GUT model  \cite{Aulakh:1982sw,Aulakh:2003kg} made use of a four-index antisymmetric tensor $210_{H}$ instead of $45_{H}$. Moreover, $210_{H}$ provided a renormalizable coupling between $126_{H}$ and $10_{H}$ (the minimal set of Higgses that can support a potentially realistic flavour structure in renormalizable $SO(10)$ GUTs) which is necessary for a correct ``distribution'' of the electroweak VEV among the two multiplets. 
Though this setting was successful in accommodating all the flavour data and even provided an interesting argument in favour of a relatively large leptonic 1-3 mixing angle (which typically fell within the range reported by the recent first direct measurements by Double-Chooz~\cite{Abe:2011fz}, Daya-Bay~\cite{An:2012eh} and RENO~\cite{Ahn:2012nd}), the ``minimal SUSY GUT'' did not pass the full consistency tests \cite{Aulakh:2005mw,Bertolini:2006pe} revealing a strong tension between phenomenological bounds on the absolute neutrino masses and the gauge unification and proton decay constraints. 

\paragraph{Quantum salvation of the minimal non-supersymmetric $SO(10)$ Higgs model}  

Remarkably enough, the above mentioned fatal vacuum instability emerging in the non-supersymmetric  $45_{H}+16_{H}/126_{H}$ SO(10) Higgs model have been recently~\cite{Bertolini:2009es} shown to be a mere artefact of the tree-level analysis used in~\cite{Buccella:1980qb,Yasue:1980fy,Anastaze:1983zk,Babu:1984mz} and there turns out to be no such a no-go at the level of the full one-loop effective potential. In particular, it has been shown that there exist stable vacua that can support two vastly different VEVs of the adjoint $45_{H}$ that avoids the need to pass through the problematic intermediate $SU(5)$ stage.

\paragraph{The persistent absolute neutrino mass scale issue}In spite of this success there still remains a certain degree of tension between the absolute neutrino mass scale and the gauge coupling unification in the model under consideration. Indeed, the general gauge unification studies~\cite{Chang:1984qr,Deshpande:1992em,Bertolini:2009qj} suggest that the scale of the right-handed neutrino masses (which in the renormalizable settings with $126_{H}$ corresponds -- up to the relevant Yukawa couplings -- to the $B-L$ symmetry breaking scale) does not exceed roughly $10^{10}$ GeV. This, however, points to light neutrino masses well above the bounds imposed by cosmology and neutrinoless double-beta decay. 

\paragraph{Reconciling the absolute neutrino mass scale with gauge unification constraints}An obvious way out may consist in tuning the entire neutrino Dirac Yukawa matrix down by several orders of magnitude. However, this would amount to a number of additional constraints on the flavour structure of the theory which is already rather rigid due to its $SO(10)$  nature. Alternatively~\cite{Bertolini:2012im}, one can attempt to push the $B-L$ breaking scale up to the required  $10^{13\div 14}$ GeV domain by ``populating'' the desert by a suitable multiplet\footnote{Since we consider only models with a minimal and chiral matter contents such a multiplet must be supplied by the scalar sector of the theory.} which can take over the role of a ``non-SUSY running helper'' and, thus, lift the usual constraints the gauge unification studies~\cite{Chang:1984qr,Deshpande:1992em,Bertolini:2009qj} imposed on the $B-L$ sector (i.e., on the $B-L$-breaking Higgs multiplet and the relevant gauge boson). This is the scenario we shall focus on here. 
\section{Consistent gauge unification in the $45_{H}+126_{H}$ $SO(10)$ Higgs model}
In a full theory, pushing a chosen multiplet away from its natural\footnote{By ``natural'' we mean the typical scale at which the multiplet's mass is expected according to the ``minimal survival hypothesis''~\cite{delAguila:1980at}.} scale does not need to be entirely trivial. Indeed, the entire scalar spectrum is driven by a limited number of parameters that must conspire in order to pull the desired multiplet's mass down to the GUT desert. This, however, can shift the vacuum of the theory out of the stability region and/or lower the GUT scale below the limits implied by the proton lifetime. Moreover, it is not a-priori clear which of the numerous SM sub-multiplets of $45_{H}+126_{H}$ (see, e.g., ~\cite{Bertolini:2012im}) would be most suitable for such a task. In this section we summarize the dedicated analysis of this problem performed in~\cite{Bertolini:2012im} and comment further on the prospects of a potentially realistic and predictive $SO(10)$ GUT based on the Higgs sector under consideration. 
\subsection{The renormalizable $45_{H}+126_{H}$ $SO(10)$ Higgs model}  
Adopting the notation and conventions of \cite{Bertolini:2012im} the scalar potential of the renormalizable 45+126 $SO(10)$ Higgs model can be written as
$
V = V_{45} + V_{126} + V_{\rm mix}
$
where
\begin{eqnarray}\label{potential}
 V_{45} &=& - \frac{\mu^2}{2} (\phi \phi)_0 + \frac{a_0}{4} (\phi \phi)_0 (\phi \phi)_0 + \frac{a_2}{4} (\phi \phi)_2 (\phi \phi)_2 \, , \\ 
\label{V126}
V_{126} &=&  - \frac{\nu^2}{5!} (\Sigma \Sigma^*)_0\!+\! \frac{\lambda_0}{(5!)^2} (\Sigma \Sigma^*)_0 (\Sigma \Sigma^*)_0 
 \!+\! \frac{\lambda_2}{(4!)^2} (\Sigma \Sigma^*)_2 (\Sigma \Sigma^*)_2 \!+\!  \frac{\lambda_4}{(3!)^2(2!)^2} (\Sigma \Sigma^*)_4 (\Sigma \Sigma^*)_4
 \!+\! \frac{\lambda'_{4}}{(3!)^2} (\Sigma \Sigma^*)_{4'} (\Sigma \Sigma^*)_{4'} \nonumber 
 \\ & &
 + \left[\frac{\eta_2}{(4!)^2} (\Sigma \Sigma)_2 (\Sigma \Sigma)_2+h.c.\right]
\nonumber \\ 
\label{V45126}
V_{\rm mix} &=& \frac{i \tau}{4!} (\phi)_2 (\Sigma \Sigma^*)_2 
+ \frac{\alpha}{2 \cdot 5!} (\phi \phi)_0 (\Sigma \Sigma^*)_0 + \frac{\beta_4}{4 \cdot 3!} (\phi \phi)_4 (\Sigma \Sigma^*)_4
+ \frac{\beta'_{4}}{3!} (\phi \phi)_{4'} (\Sigma \Sigma^*)_{4'} \nonumber 
+ \left[\frac{\gamma_2}{4!} (\phi \phi)_2 (\Sigma \Sigma)_2+ h.c.\right]
\end{eqnarray}
\begin{eqnarray}
\text{and
}& (\phi \phi)_0  \equiv \phi_{ij} \phi_{ij}, \;\; (\Sigma \Sigma^*)_0 \equiv \Sigma_{ijklm} \Sigma^*_{ijklm}\,,\;\;  
 (\phi \phi)_0 (\phi \phi)_0  \equiv \phi_{ij} \phi_{ij} \phi_{kl} \phi_{kl} \,,\;\;   (\phi \phi)_2 (\phi \phi)_2  \equiv \phi_{ij} \phi_{ik} \phi_{lj} \phi_{lk} \,, & \\
& (\Sigma \Sigma^*)_0 (\Sigma \Sigma^*)_0  \equiv \Sigma_{ijklm} \Sigma^*_{ijklm} \Sigma_{nopqr} \Sigma^*_{nopqr}  \,,\;\;   (\Sigma \Sigma^*)_2 (\Sigma \Sigma^*)_2  \equiv \Sigma_{ijklm} \Sigma^*_{ijkln} \Sigma_{opqrm} \Sigma^*_{opqrn}\,,  & \nonumber \\
& (\Sigma \Sigma^*)_4 (\Sigma \Sigma^*)_4  \equiv \Sigma_{ijklm} \Sigma^*_{ijkno} \Sigma_{pqrlm} \Sigma^*_{pqrno} \,,\;\;   (\Sigma \Sigma^*)_{4'} (\Sigma \Sigma^*)_{4'}  \equiv \Sigma_{ijklm} \Sigma^*_{ijkno} \Sigma_{pqrln} \Sigma^*_{pqrmo}\,,   &\nonumber \\
& (\Sigma \Sigma)_2 (\Sigma \Sigma)_2  \equiv \Sigma_{ijklm} \Sigma_{ijkln} \Sigma_{opqrm} \Sigma_{opqrn}   \,,\;\;   (\Sigma^* \Sigma^*)_2 (\Sigma^* \Sigma^*)_2  \equiv \Sigma^*_{ijklm} \Sigma^*_{ijkln} \Sigma^*_{opqrm} \Sigma^*_{opqrn}\,,  & \nonumber \\
& (\phi)_2 (\Sigma \Sigma^*)_2 \equiv \phi_{ij} \Sigma_{klmni} \Sigma^*_{klmnj}   \,,\;\;   (\phi \phi)_0 (\Sigma \Sigma^*)_0 \equiv \phi_{ij} \phi_{ij} \Sigma_{klmno} \Sigma^*_{klmno}\,, \;\;\;   (\phi \phi)_4 (\Sigma \Sigma^*)_4 \equiv \phi_{ij} \phi_{kl} \Sigma_{mnoij} \Sigma^*_{mnokl}\,, & \nonumber\\
&  (\phi \phi)_{4'} (\Sigma \Sigma^*)_{4'} \equiv \phi_{ij} \phi_{kl} \Sigma_{mnoik} \Sigma^*_{mnojl}\,,\;\;\; (\phi \phi)_2 (\Sigma \Sigma)_2 \equiv \phi_{ij} \phi_{ik} \Sigma_{lmnoj} \Sigma_{lmnok}  \,,\;\;   (\phi \phi)_2 (\Sigma^* \Sigma^*)_2 \equiv \phi_{ij} \phi_{ik} \Sigma^*_{lmnoj} \Sigma^*_{lmnok}   &\nonumber
\end{eqnarray}
is the maximal set of linearly independent renormalizable invariants one can construct out of the components $\phi_{ij}$ of the adjoint representation and the components $\Sigma_{ijklm}$ of the self-dual part of the fully antisymmetric five-index tensor of $SO(10)$. The scalar spectrum resulting from the minimization of $V$ is given in Appendix B of \cite{Bertolini:2012im}. 

\paragraph{Parameter counting}The scalar potential (\ref{potential}) is driven in total by 14 real parameters and 2 phases. 
Let us also recall that there are three SM singlet VEVs residing
in the $(1,1,1,0)$ and $(1,1,3,0)$ sub-multiplets of $45_{H}$
and in the $(1,1,3,+2)$ component of $126_{H}$ where the quantum numbers correspond to the $SU(3)_{c}\times SU(2)_{L}\times SU(2)_{R}\times U(1)_{B-L}$ subgroup of $SO(10)$. The minimization conditions constrain three combinations of the parameters above (we choose $\mu$, $\nu$ and $a_{2}$). Since, however, $\eta$ does not enter the scalar masses at all and $\gamma_{2}$ enters only as $|\gamma_{2}|^{2}$ one is in general left with just 10 real numbers parametrizing the scalar spectrum\footnote{For further details see the discussion in~\cite{Bertolini:2012im}. Note, for instance, that ${\rm Arg}(\gamma)$ should in general pop up in the diagonalization matrices.}. Finally, an attempt to reconcile the seesaw scale with the light neutrino mass limits by pushing one extra scalar into the desert amounts to one (and only one!) more fine-tuning, so for each choice of the three VEVs constrained by gauge unification and proton decay, one is left with only 9 independent real parameters, cf. TABLE~\ref{TableSampleParameters}. 

\paragraph{Predictivity prospects}In connection to this, let us recall that the predictive power of the minimal SUSY $SO(10)$ was related mainly the fact that its SM vacuum manifold was parametrized by only two real numbers~\cite{Bajc:2004xe}. In this respect, the minimal non-SUSY $SO(10)$ scenario seems to be in a much worse shape.  Nevertheless, it is rather premature to claim that the model is free of any interesting and robust low-energy predictions. Indeed, it is clear that not all  configurations of the 9 independent parameters correspond to stable vacua of the theory\footnote{Actually, a homogeneous random scan over the parameter space reveals that, in most cases, they do not. In this respect, the expectation that vacuum stability can play a decisive role in the predictive power of the current setting is not entirely unsupported.}. Thus, the real predictive power of the model under consideration remains unclear and should be studied in detail in convolution with all the other available phenomenology constraints (i.e., flavour structure, proton lifetime, Big-Bang nucleosynthesis). In this respect one should also not forget about the need for an extra $10_{H}$ in the Higgs sector in order to make the Yukawa structure potentially realistic. Nevertheless, since $10_{H}$ does not participate at the high-scale symmetry breaking the vacuum stability considerations remain unaffected.  

However, such a complete analysis, though highly desirable, was beyond the scope of the study~\cite{Bertolini:2012im} and will be elaborated elsewhere. Here we shall focus solely on the first step in the desired direction, i.e., on the identification of the settings in which the seesaw scale falls  into the $10^{13\div 14}$~GeV ballpark and, hence, provides the best chance to accommodate the light neutrino masses within the relevant experimental limits.

\subsection{Basic consistency constraints}  

\paragraph{Non-tachyonic scalar spectrum \& radiative corrections}  

In doing so, one has to address the question of the radiative stabilization of the physically interesting vacua of the current model. Unlike in the $45_{H}+16_{H}$ case that has been worked out in great detail in reference~\cite{Bertolini:2009es}, the full-fledged calculation of the one-loop effective potential is very difficult with $45_{H}+126_{H}$ in the Higgs sector due to the higher number and complexity of the $\Sigma$-invariants in the tree-level potential above. 

Nevertheless, as it was argued in~\cite{Bertolini:2012im} it is (at least for the time being) sufficient to focus at the most universal scalar one-loop correction, namely, at the leading non-logarithmic $SO(10)$-invariant $\tau^{2}$ term which can be evaluated by a simple diagrammatic calculation. Indeed, as it yields a positive correction to all scalar masses, it should  be enough to regularize the key tachyonic instabilities of the tree-level spectrum\footnote{As in~\cite{Bertolini:2009es} we expect the other leading non-logarithmic corrections to be positive and, hence, including just the $SO(10)$ invariant piece can be viewed as a minimal way to stabilize the tachyons.}, c.f.~\cite{Bertolini:2009es}.
It is not difficult to see that such leading scalar-loop induced non-logarithmic corrections come from tadpole type of diagrams and that the $\tau^{2}$-proportional non-logarithmic term emerges only through the renormalization of the stationary conditions for the $45_{H}$ VEVs. Diagrammatically, they correspond to the one-loop correction to the one-point function of $45_{H}$ of the form 
\begin{equation}
\parbox{2.4cm}{\includegraphics[width=2.4cm]{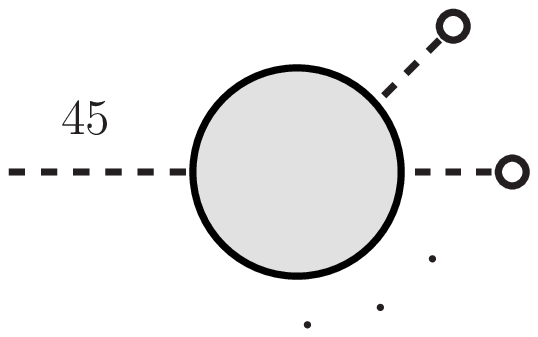}}\ni
\parbox{2.6cm}{\includegraphics[width=2.6cm]{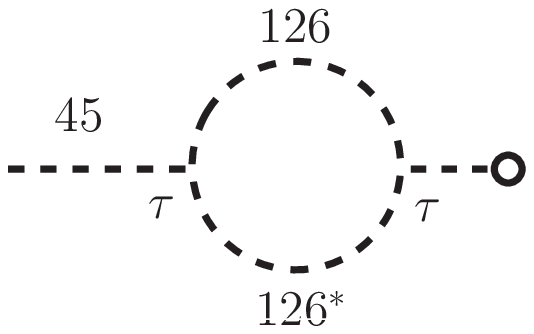}}+
\text{higher order terms}
\end{equation}
where the ``higher order terms'' denote diagrams with higher number of VEV  insertions (via the $\tau$-vertex).
Given the $SO(10)$ structure of the relevant $\tau$-term in~(\ref{potential})
the universal mass shift due to this class of graphs reads
\begin{equation}\label{leadingoneloopcorrection}
\Delta M^{2}_{\text{1-loop-}\tau^{2}}=\frac{35\tau^{2}}{32\pi^{2}}+\text{logs}\,,
\end{equation}
where the symbol ``logs'' denotes all the logarithmic corrections that are minimized at  the GUT scale. 

\paragraph{Proton decay and Big Bang nucleosynthesis}  
Without a detailed information about the flavour structure of a specific model the best one can do about the $d=6$ proton decay limits is to focus on the ``universal'' gauge transitions parametrized (in the current basis) by the unified gauge coupling and the GUT  breaking scale $M_{G}$ and assume ${\cal O}(1)$ entries of the unitary matrices corresponding to the rotations from the current to the mass bases; in such a case the position of the gauge unification point translates directly to the proton lifetime. For that, we shall impose namely the latest (2011) Super-Kamiokande limit~\cite{Nakamura:2010zzi}:
\begin{equation}\label{limit:SK}
\tau(p\to e^{+}\pi^{0})_{\rm SK, 2011}> 8.2 \times 10^{33}\, {\rm years}\,,
\end{equation}
and, for sake of illustration, a couple of assumed future sensitivity limits that 
Hyper-Kamiokande (HK)~\cite{Abe:2011ts} should reach by 2025 and 2040:
\begin{equation}
\label{limit:HK}\tau(p\to e^{+}\pi^{0})_{\rm HK, 2025} >   9 \times 10^{34}\, {\rm years}\,,\qquad
\tau(p\to e^{+}\pi^{0})_{\rm HK, 2040}  >  2 \times 10^{35}\, {\rm years}\,.
\end{equation}
On the Big Bang nucleosynthesis (BBN) side, we require that any accidentally light scalar coupled to the SM matter decays well before the BBN phase of the early Universe evolution in order to prevent any distortion of the successful standard BBN picture~\cite{Nakamura:2010zzi}. We do not expect any trouble here because all the candidate multiplets identified in~\cite{Bertolini:2012im} come from $126_{H}$ whose  renormalizable Yukawa coupling to the SM matter are assumed to be non-negligible. 
\paragraph{Gauge unification with an accidentally light scalar multiplet in the desert}  
Last, but not least, we require that the heavy scalar spectrum governed by the 9 parameters identified above (see also TABLE~\ref{TableSampleParameters}) is such that the SM gauge couplings properly unify at high energies and that the GUT-scale gauge spectrum (driven by the dominant $SO(10)$-breaking VEV and the relevant gauge coupling) is compatible with the current proton-lifetime constraint~(\ref{limit:SK}). Technically, it is convenient to work with the three effective SM gauge couplings rather than redefine the set of gauge parameters whenever a restoration of a higher intermediate gauge symmetry is encountered in the gauge and scalar spectra (i.e, whenever the relevant VEV is surpassed). 

One should also note that the results obtained in~\cite{Bertolini:2012im} are subject to several sources of theoretical uncertainties. In particular, given the need for a radiative stabilization of the tree-level  tachyons, a full-fledged two-loop renormalization group analysis is highly desirable. In this respect, the one-loop running performed in~\cite{Bertolini:2012im} should be interpreted as an indication of viability rather than a full-fledged analysis; nevertheless, even such a simplified picture should account for all the qualitative features of the full two-loop approach and, as such, it can be viewed as a good first approximation. For a more detailed discussion an interested reader is deferred to the original work~\cite{Bertolini:2012im}.
\subsection{Upper bounds on the seesaw scale in the minimal $SO(10)$ Higgs model revisited}
\begin{figure}[th]
 \includegraphics[width=8cm, height=5.3cm]{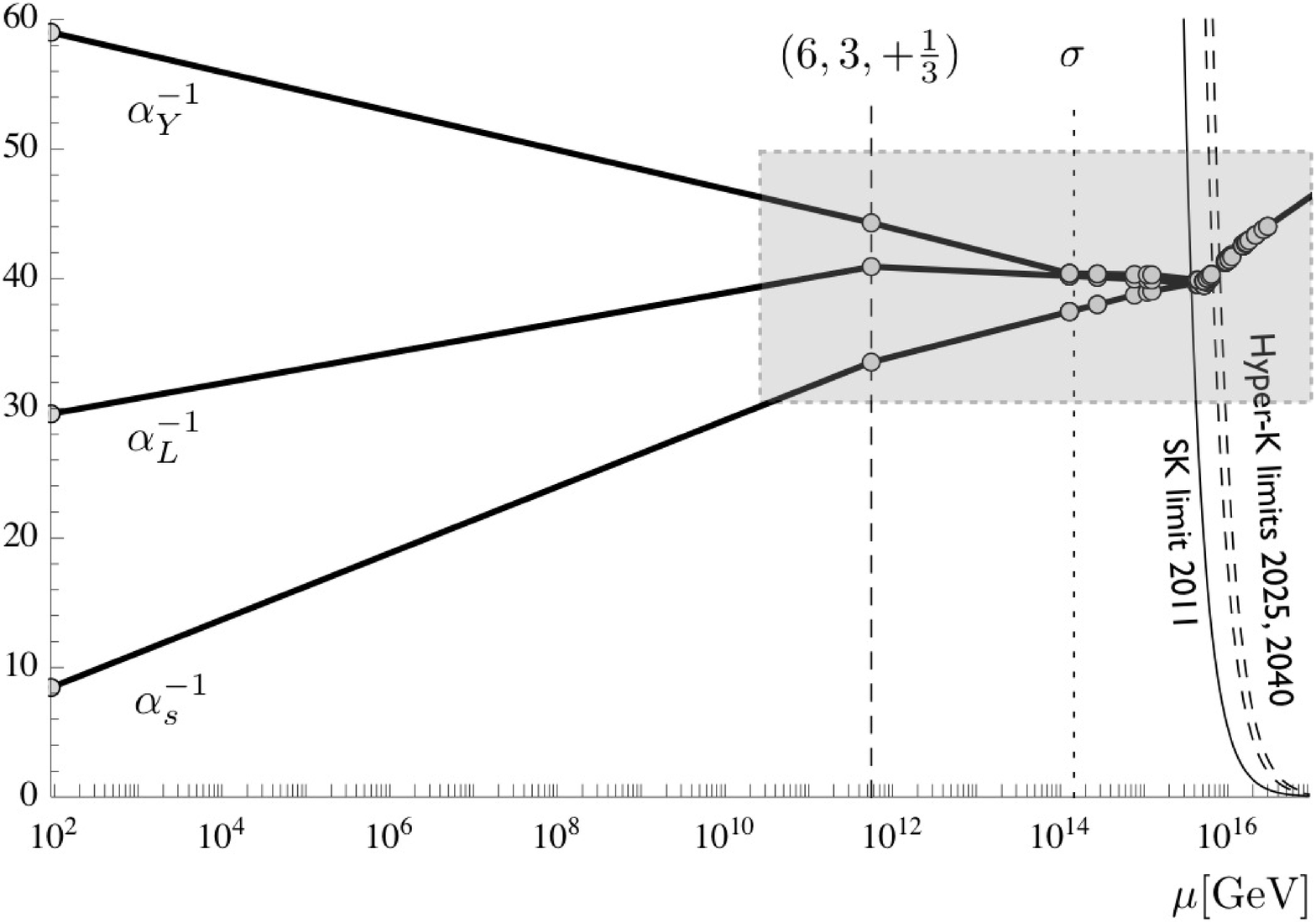}
 \raisebox{3.5mm}{\includegraphics[width=8cm, height=5.1cm]{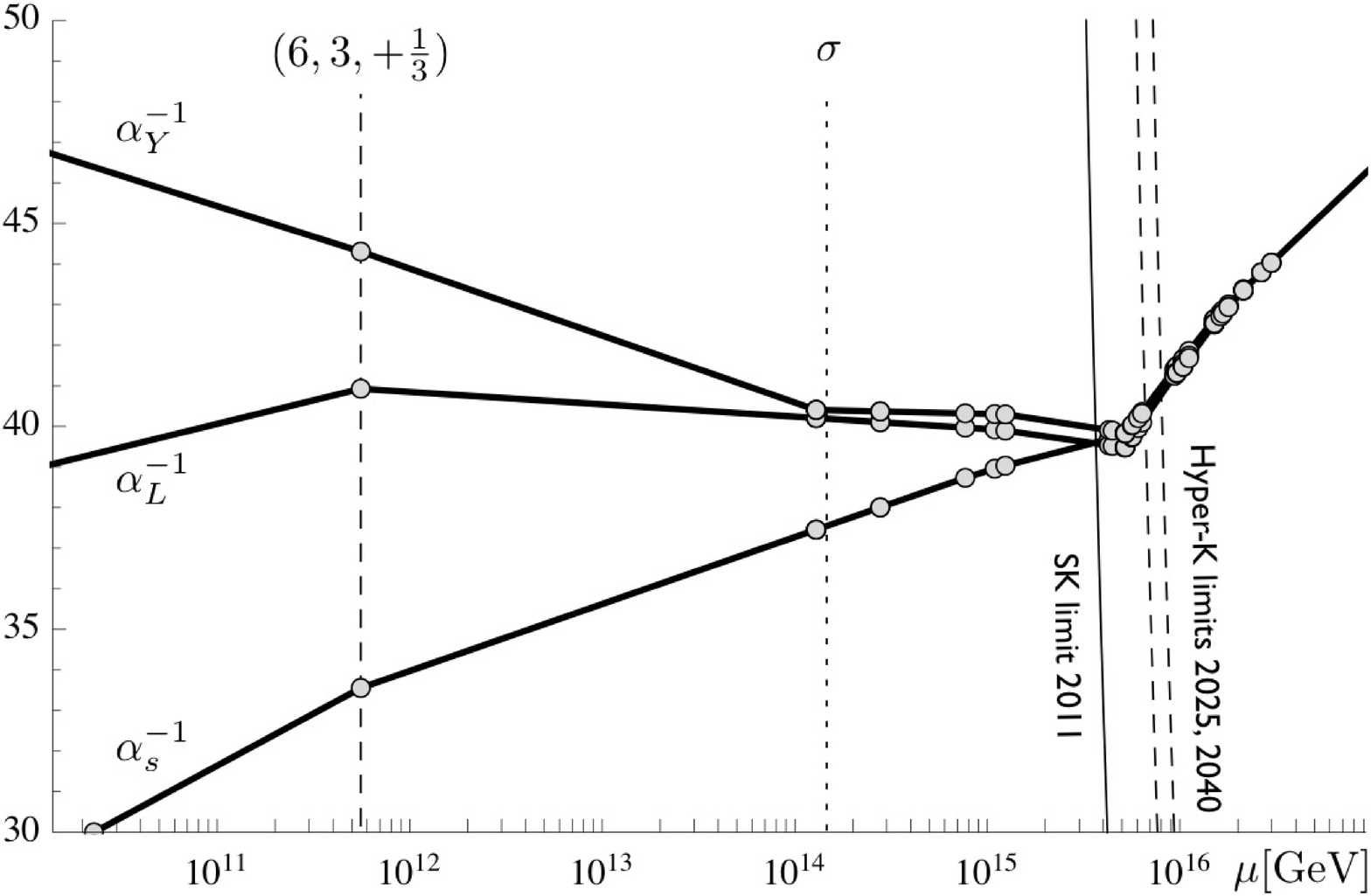}}
 \caption{\label{Figure1}Unification of the effective SM gauge couplings in the sample setting with a light $(6,3,+\tfrac{1}{3})$ multiplet (here at around $5.6\times 10^{11}$GeV, cf. TABLE~\ref{TableSampleParameters}) with the shaded area magnified in  the right panel. The small circles indicate the positions of various thresholds (for details, see TABLE~\ref{TableSpectra}) inflicting changes in the three curve's slopes. The almost-vertical solid and dashed lines correspond to the current Super-Kamiokande and assumed future Hyper-Kamiokande proton lifetime limits~(\ref{limit:SK}), (\ref{limit:HK}). The dotted vertical line indicates the position of the $B-L$ breaking scale at about $10^{14}$~GeV.}
\end{figure}

\begin{figure}[h]
  \includegraphics[width=8cm, height=5.3cm]{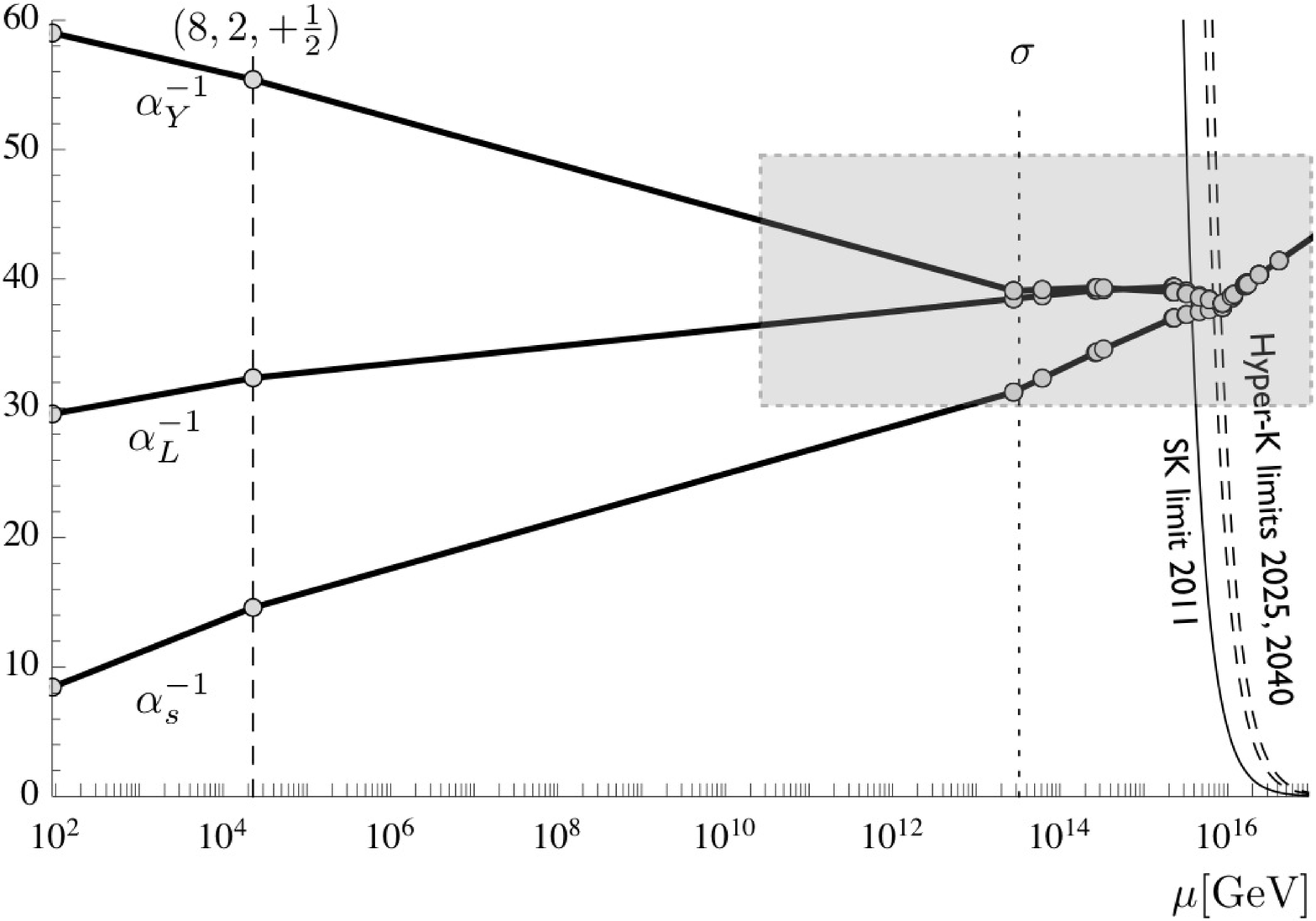} 
  \raisebox{3mm}{\includegraphics[width=8cm, height=5.1cm]{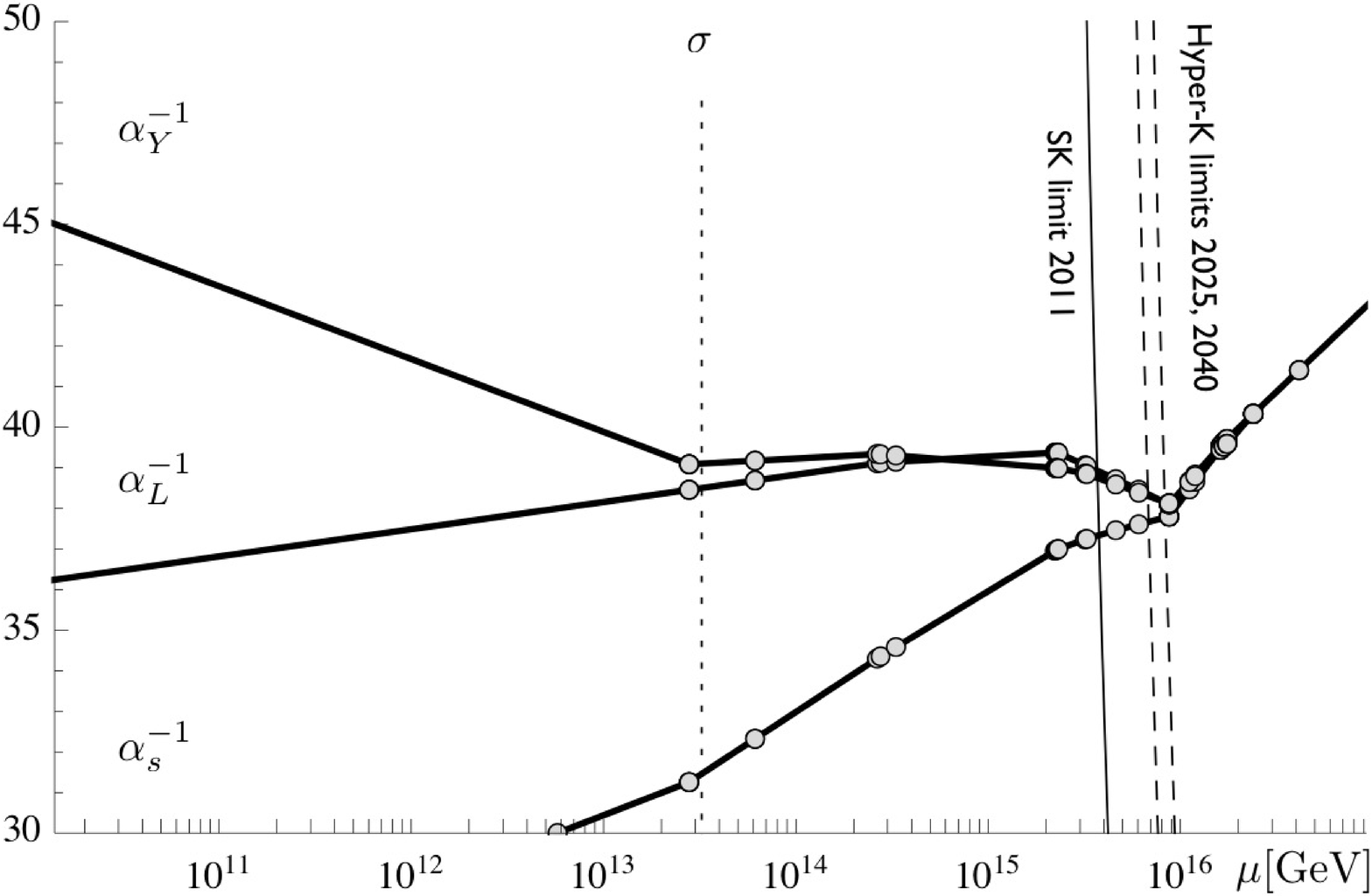}}
  \caption{\label{Figure2}The same as in FIGURE~\ref{Figure1} but this time for a light $(8,2,+\tfrac{1}{2})$ at around $2.3\times 10^{4}$~GeV. The $B-L$ breaking scale indicated by the dotted vertical line is again shifted from the naive upper bound~\cite{Bertolini:2009qj} (at around $10^{10}$~GeV) to the vicinity of $10^{14}$~GeV. The masses of the relevant GUT- and intermediate-scale thresholds are listed in TABLE~\ref{TableSpectra}.}
\end{figure}

With an extra light multiplet pulled into the desert the generic upper bounds on the $B-L$ breaking scale $M_{BL}$ obtained in~\cite{Bertolini:2009qj} under the minimal survival hypothesis are modified substantially. There are two qualitatively different regions of the parameter space identified in~\cite{Bertolini:2012im} where $M_{BL}$ can be pushed up to the favourable level of $10^{13\div 14}$~GeV and, at the same time, vacuum stability and a full compatibility with namely the proton decay and BBN constraints is maintained. In one case,  a light $(6,3,+\tfrac{1}{3})$ pulled down to about $10^{11}$ GeV reshapes the gauge unification pattern so that the $B-L$ scale as high as $10^{14}$~GeV becomes easily accessible. In the other case, a light multiplet transforming as $(8,2,+\tfrac{1}{2})$ of the SM in the lower part of the GUT desert (i.e., below $10^{8}$~GeV) provides enough room for the $B-L$ scale to be shifted up to about $10^{13}$~GeV.
Specific numerical examples of these two basic settings are given in TABLE~\ref{TableSampleParameters}; the relevant scalar spectra are listed in TABLE~\ref{TableSpectra} and the resulting gauge unification patterns are depicted in FIGUREs~\ref{Figure1} and \ref{Figure2}.
\begin{table}[t]
\hskip 2.5cm
\begin{tabular}{c|r}
\hline
$M(6,3,+\tfrac{1}{3})$ [GeV] & $5.6\times 10^{11}$  \\
\hline
$\omega_{R}$ [GeV] & $-2.92 \times 10^{13}$    \\
$\omega_{BL}$ [GeV] & ${\bf 8.65  \times 10^{15}}$    \\
$\sigma$ [GeV] &  $-1.46  \times 10^{14}$  \\
$a_{0}$ &  $ 0.50 $   \\
$\alpha$ &  $ 0.55 $   \\
$\beta_{4} $ &  $ 0.61 $ \\
$\beta_{4}'$ &  $ -0.41 $  \\
$|\gamma_{2}|$ &  $ 0.12 $  \\
$\lambda_{0}$ &  $ 0.95 $  \\
$\lambda_{2}$ &  $ 0.34  $ \\
$\lambda_{4}$ &  $ -0.07  $\\
$\lambda_{4}' $ &  $ -0.15  $ \\
\hline
\end{tabular}
\hskip 1cm
\begin{tabular}{c|r}
\hline
$M(8,2,+\tfrac{1}{2})$ [GeV]  & $2.3 \times 10^{4}$ \\
\hline
 $\omega_{R}$ [GeV] & ${\bf -1.46 \times 10^{16}}$   \\
$\omega_{BL}$ [GeV]   & $-4.04 \times 10^{12}$  \\
$\sigma$ [GeV]  & $-3.23  \times 10^{13}$ \\
$a_{0}$  & $ 0.50$  \\
$\alpha$  & $ 0.47 $  \\
$\beta_{4} $  & $0.60 $ \\
$\beta_{4}'$   & $ -0.34 $\\
$|\gamma_{2}|$   & $  0.01 $\\
$\lambda_{0}$  & $ -0.86 $ \\
$\lambda_{2}$ & $ -0.14 $ \\
$\lambda_{4}$ & $ -0.04$\\
$\lambda_{4}' $  & $ -0.07$\\
\hline
\end{tabular}
\hskip 2.5cm\mbox{}
\caption{\label{TableSampleParameters}Parameters underpinning the two sample settings depicted in FIGUREs~\ref{Figure1} (left) and \ref{Figure2} (right), see also TABLE~\ref{TableSpectra}. The value of $\tau$ is determined from the mass of  the accidentally light multiplet given in the first row. The largest VEV (i.e., the GUT symmetry breaking scale) is in boldface.}
\end{table}
\section{Conclusions and outlook}
The non-supersymmetric $SO(10)$ Higgs model with the $45$- and $126$-dimensional multiplets responsible for the breaking of the GUT-scale gauge symmetry down to the SM gauge group has been widely ignored for several decades due to a generic tree-level vacuum stability issue encountered within all potentially realistic gauge unification patterns. With the recent observation that radiative corrections can resolve this problem the idea that a renormalizable model based on this setting can account for a calculable and potentially realistic $SO(10)$ GUT was put forward in~\cite{Bertolini:2012im}. In this article we summarized all the basic ingredients of the model paying particular attention to the vacuum stability, gauge unification and proton lifetime constraints and commented in brief on the prospects of a future complete analysis including flavour. In particular, it was shown that there is enough room for a $B-L$ breaking scale as high as $10^{13\div 14}$~GeV, right in the region favoured by the seesaw picture of the neutrino mass generation. 

Nevertheless, at the current level, the study~\cite{Bertolini:2012im} should be taken only as a promising indication of viability rather than a complete feasibility check of the framework under consideration. Indeed, besides a full-fledged two-loop gauge unification analysis (which, in principle, requires a full one-loop information about the scalar spectrum of the theory), such an ultimate test  requires namely a thorough inspection of the proton decay constraints which, however, can not be done without a detailed flavour sector analysis.     

\begin{theacknowledgments}
M.M. would like to thank the organizers of the GUT'2012 workshop for the warm hospitality during his stay in Kyoto. S.B. is associated to the theoretical particle physics group at SISSA, Trieste. The work of L.DL. was supported by the DFG through the SFB/TR 
9 ``Computational Particle Physics''. The work of M.M. is supported by the Marie Curie Intra European Fellowship
within the 7th European Community Framework Programme
FP7-PEOPLE-2009-IEF, contract  PIEF-GA-2009-253119, by the EU
Network grant UNILHC PITN-GA-2009-237920, by the Spanish MICINN
grants FPA2008-00319/FPA and MULTIDARK CAD2009-00064
(Consolider-Ingenio 2010 Programme) and by the Generalitat
Valenciana grant Prometeo/2009/091.
\end{theacknowledgments}

\appendix
\renewcommand{\arraystretch}{1.3}
\begin{table}
\begin{tabular}{c|c|c|c}
\hline
multiplet & type  & $\Delta b^{321}$ & mass [GeV]  \\
\hline
$\bf (6,3,+\tfrac{1}{3})$ & {\bf CS} & $(\tfrac{5}{2}, 4, \tfrac{2}{5})$ & $\bf 5.6 \times 10^{11}$ \\
$(1,1,-1)$ & VB & $(0, 0, -\tfrac{11}{5})$ & $1.3 \times 10^{14}$ \\
$(1,1,+1)$ & VB & $(0, 0, -\tfrac{11}{5})$ & $1.3 \times 10^{14}$ \\
$(1,1,+1)$ & GB & $(0, 0, \tfrac{1}{5})$ & $1.3 \times 10^{14}$ \\
$(1,1,0)$ & VB & $(0, 0, 0)$ & $2.8 \times 10^{14}$ \\
$(1,1,0)$ & GB & $(0, 0, 0)$ & $2.8 \times 10^{14}$ \\
$(8,1,0)$ & RS & $(\tfrac{1}{2}, 0, 0)$ & $7.7 \times 10^{14}$ \\
$(3,2,+\tfrac{1}{6})$ & CS & $(\tfrac{1}{3}, \tfrac{1}{2}, \tfrac{1}{30})$ & $1.1 \times 10^{15}$ \\
$(3,2,+\tfrac{7}{6})$ & CS & $(\tfrac{1}{3}, \tfrac{1}{2}, \tfrac{49}{30})$ & $1.2 \times 10^{15}$ \\
$(1,1,0)$ & RS & $(0, 0, 0)$ & $4.3 \times 10^{15}$ \\
$(1,1,+2)$ & CS & $(0, 0, \tfrac{4}{5})$ & $4.5 \times 10^{15}$ \\
$\bf(\overline{3},2,-\tfrac{1}{6})$ & \bf VB & $(-\tfrac{11}{3}, -\tfrac{11}{2}, -\tfrac{11}{30})$ & $\bf5.2 \times 10^{15}$ \\
$\bf(3,2,+\tfrac{1}{6})$ & \bf VB & $(-\tfrac{11}{3}, -\tfrac{11}{2}, -\tfrac{11}{30})$ & $\bf5.2 \times 10^{15}$ \\
$\bf (3,2,+\tfrac{1}{6})$ & \bf GB & $(\tfrac{1}{3}, \tfrac{1}{2}, \tfrac{1}{30})$ & $\bf 5.2 \times 10^{15}$ \\
$\bf(\overline{3},2,+\tfrac{5}{6})$ & \bf VB & $(-\tfrac{11}{3}, -\tfrac{11}{2}, -\tfrac{55}{6})$ & $\bf5.2 \times 10^{15}$ \\
$\bf(3,2,-\tfrac{5}{6})$ & \bf VB & $(-\tfrac{11}{3}, -\tfrac{11}{2}, -\tfrac{55}{6})$ & $\bf5.2 \times 10^{15}$ \\
$\bf (3,2,-\tfrac{5}{6})$ & \bf GB & $(\tfrac{1}{3}, \tfrac{1}{2}, \tfrac{5}{6})$ & $\bf 5.2 \times 10^{15}$ \\
$(1,1,+1)$ & CS & $(0, 0, \tfrac{1}{5})$ & $5.6 \times 10^{15}$ \\
$(1,1,0)$ & RS & $(0, 0, 0)$ & $5.7 \times 10^{15}$ \\
$(1,3,0)$ & RS & $(0, \tfrac{1}{3}, 0)$ & $6.1 \times 10^{15}$ \\
$(\overline{3},1,+\tfrac{1}{3})$ & CS & $(\tfrac{1}{6}, 0, \tfrac{1}{15})$ & $6.4 \times 10^{15}$ \\
$(8,2,+\tfrac{1}{2})$ & CS & $(2, \tfrac{4}{3}, \tfrac{4}{5})$ & $9.3 \times 10^{15}$ \\
$(\overline{3},1,+\tfrac{4}{3})$ & CS & $(\tfrac{1}{6}, 0, \tfrac{16}{15})$ & $9.6 \times 10^{15}$ \\
$(\overline{3},1,+\tfrac{1}{3})$ & CS & $(\tfrac{1}{6}, 0, \tfrac{1}{15})$ & $9.6 \times 10^{15}$ \\
$(\overline{3},1,-\tfrac{2}{3})$ & CS & $(\tfrac{1}{6}, 0, \tfrac{4}{15})$ & $9.6 \times 10^{15}$ \\
$(\overline{3},1,-\tfrac{2}{3})$ & VB & $(-\tfrac{11}{6}, 0, -\tfrac{44}{15})$ & $1.0 \times 10^{16}$ \\
$(3,1,+\tfrac{2}{3})$ & VB & $(-\tfrac{11}{6}, 0, -\tfrac{44}{15})$ & $1.0 \times 10^{16}$ \\
$(\overline{3},1,-\tfrac{2}{3})$ & GB & $(\tfrac{1}{6}, 0, \tfrac{4}{15})$ & $1.0 \times 10^{16}$ \\
$(8,2,+\tfrac{1}{2})$ & CS & $(2, \tfrac{4}{3}, \tfrac{4}{5})$ & $1.1 \times 10^{16}$ \\
$(\overline{6},1,+\tfrac{2}{3})$ & CS & $(\tfrac{5}{6}, 0, \tfrac{8}{15})$ & $1.5 \times 10^{16}$ \\
$(1,2,+\tfrac{1}{2})$ & {RS} & $(0, \tfrac{1}{12}, \tfrac{1}{20})$ & $1.5 \times 10^{16}$ \\
$(\overline{6},1,-\tfrac{1}{3})$ & CS & $(\tfrac{5}{6}, 0, \tfrac{2}{15})$ & $1.5 \times 10^{16}$ \\
$(\overline{6},1,-\tfrac{4}{3})$ & CS & $(\tfrac{5}{6}, 0, \tfrac{32}{15})$ & $1.5 \times 10^{16}$ \\
$(1,2,+\tfrac{1}{2})$ & {RS} & $(0, \tfrac{1}{12}, \tfrac{1}{20})$ & $1.6 \times 10^{16}$ \\
$(\overline{3},1,+\tfrac{1}{3})$ & CS & $(\tfrac{1}{6}, 0, \tfrac{1}{15})$ & $1.7 \times 10^{16}$ \\
$(3,3,-\tfrac{1}{3})$ & CS & $(\tfrac{1}{2}, 2, \tfrac{1}{5})$ & $1.8 \times 10^{16}$ \\
$(3,2,+\tfrac{1}{6})$ & CS & $(\tfrac{1}{3}, \tfrac{1}{2}, \tfrac{1}{30})$ & $2.1 \times 10^{16}$ \\
$(3,2,+\tfrac{7}{6})$ & CS & $(\tfrac{1}{3}, \tfrac{1}{2}, \tfrac{49}{30})$ & $2.1 \times 10^{16}$ \\
$(1,3,-1)$ & CS & $(0, \tfrac{2}{3}, \tfrac{3}{5})$ & $2.6 \times 10^{16}$ \\
$(1,1,0)$ & RS & $(0, 0, 0)$ & $3.0 \times 10^{16}$ \\
\hline
\end{tabular}
\;\;\;\;\;\;
\begin{tabular}{c|c|c|c}
\hline
multiplet & type   & $\Delta b^{321}$ & mass [GeV]  \\
\hline
$\bf (8,2,+\tfrac{1}{2})$ & {\bf CS} & $(2, \tfrac{4}{3}, \tfrac{4}{5})$ & $\bf 2.3 \times 10^{4}$ \\
$(\overline{3},1,-\tfrac{2}{3})$ & VB & $(-\tfrac{11}{6}, 0, -\tfrac{44}{15})$ & $2.8 \times 10^{13}$ \\
$(3,1,+\tfrac{2}{3})$ & VB & $(-\tfrac{11}{6}, 0, -\tfrac{44}{15})$ & $2.8 \times 10^{13}$ \\
$(\overline{3},1,-\tfrac{2}{3})$ & GB & $(\tfrac{1}{6}, 0, \tfrac{4}{15})$ & $2.8 \times 10^{13}$ \\
$(1,1,0)$ & VB & $(0, 0, 0)$ & $6.1 \times 10^{13}$ \\
$(1,1,0)$ & GB & $(0, 0, 0)$ & $6.1 \times 10^{13}$ \\
$(3,2,+\tfrac{7}{6})$ & CS & $(\tfrac{1}{3}, \tfrac{1}{2}, \tfrac{49}{30})$ & $2.6 \times 10^{14}$ \\
$(3,2,+\tfrac{1}{6})$ & CS & $(\tfrac{1}{3}, \tfrac{1}{2}, \tfrac{1}{30})$ & $2.8 \times 10^{14}$ \\
$(1,2,+\tfrac{1}{2})$ & {RS} & $(0, \tfrac{1}{12}, \tfrac{1}{20})$ & $3.3 \times 10^{14}$ \\
$(1,1,0)$ & RS & $(0, 0, 0)$ & $2.2 \times 10^{15}$ \\
$(\overline{3},1,-\tfrac{2}{3})$ & CS & $(\tfrac{1}{6}, 0, \tfrac{4}{15})$ & $2.3 \times 10^{15}$ \\
$(6,3,+\tfrac{1}{3})$ & CS & $(\tfrac{5}{2}, 4, \tfrac{2}{5})$ & $2.3 \times 10^{15}$ \\
$(3,3,-\tfrac{1}{3})$ & CS & $(\tfrac{1}{2}, 2, \tfrac{1}{5})$ & $2.3 \times 10^{15}$ \\
$(1,3,-1)$ & CS & $(0, \tfrac{2}{3}, \tfrac{3}{5})$ & $2.3 \times 10^{15}$ \\
$(\overline{6},1,-\tfrac{4}{3})$ & CS & $(\tfrac{5}{6}, 0, \tfrac{32}{15})$ & $3.2 \times 10^{15}$ \\
$(1,1,0)$ & RS & $(0, 0, 0)$ & $3.3 \times 10^{15}$ \\
$(8,1,0)$ & RS & $(\tfrac{1}{2}, 0, 0)$ & $4.6 \times 10^{15}$ \\
$(1,3,0)$ & RS & $(0, \tfrac{1}{3}, 0)$ & $6.1 \times 10^{15}$ \\
$\bf(\overline{3},2,+\tfrac{5}{6})$ & \bf VB & $(-\tfrac{11}{3}, -\tfrac{11}{2}, -\tfrac{55}{6})$ & $\bf8.7 \times 10^{15}$ \\
$\bf(3,2,-\tfrac{5}{6})$ & \bf VB & $(-\tfrac{11}{3}, -\tfrac{11}{2}, -\tfrac{55}{6})$ & $\bf8.7 \times 10^{15}$ \\
$\bf (3,2,-\tfrac{5}{6})$ & \bf GB & $(\tfrac{1}{3}, \tfrac{1}{2}, \tfrac{5}{6})$ & $\bf 8.7 \times 10^{15}$ \\
$\bf(\overline{3},2,-\tfrac{1}{6})$ & \bf VB & $(-\tfrac{11}{3}, -\tfrac{11}{2}, -\tfrac{11}{30})$ & $\bf8.7 \times 10^{15}$ \\
$\bf(3,2,+\tfrac{1}{6})$ & \bf VB & $(-\tfrac{11}{3}, -\tfrac{11}{2}, -\tfrac{11}{30})$ & $\bf8.7 \times 10^{15}$ \\
$\bf (3,2,+\tfrac{1}{6})$ & \bf GB & $(\tfrac{1}{3}, \tfrac{1}{2}, \tfrac{1}{30})$ & $\bf 8.7 \times 10^{15}$ \\
$(\overline{3},1,+\tfrac{1}{3})$ & CS & $(\tfrac{1}{6}, 0, \tfrac{1}{15})$ & $1.1 \times 10^{16}$ \\
$(\overline{3},1,+\tfrac{1}{3})$ & CS & $(\tfrac{1}{6}, 0, \tfrac{1}{15})$ & $1.2 \times 10^{16}$ \\
$(1,1,+1)$ & CS & $(0, 0, \tfrac{1}{5})$ & $1.6 \times 10^{16}$ \\
$(\overline{3},1,+\tfrac{1}{3})$ & CS & $(\tfrac{1}{6}, 0, \tfrac{1}{15})$ & $1.6 \times 10^{16}$ \\
$(\overline{6},1,-\tfrac{1}{3})$ & CS & $(\tfrac{5}{6}, 0, \tfrac{2}{15})$ & $1.6 \times 10^{16}$ \\
$(3,2,+\tfrac{7}{6})$ & CS & $(\tfrac{1}{3}, \tfrac{1}{2}, \tfrac{49}{30})$ & $1.7 \times 10^{16}$ \\
$(1,2,+\tfrac{1}{2})$ &{RS} & $(0, \tfrac{1}{12}, \tfrac{1}{20})$ & $1.7 \times 10^{16}$ \\
$(8,2,+\tfrac{1}{2})$ & CS & $(2, \tfrac{4}{3}, \tfrac{4}{5})$ & $1.7 \times 10^{16}$ \\
$(3,2,+\tfrac{1}{6})$ & CS & $(\tfrac{1}{3}, \tfrac{1}{2}, \tfrac{1}{30})$ & $1.7 \times 10^{16}$ \\
$(1,1,-1)$ & VB & $(0, 0, -\tfrac{11}{5})$ & $1.7 \times 10^{16}$ \\
$(1,1,+1)$ & VB & $(0, 0, -\tfrac{11}{5})$ & $1.7 \times 10^{16}$ \\
$(1,1,+1)$ & GB & $(0, 0, \tfrac{1}{5})$ & $1.7 \times 10^{16}$ \\
$(1,1,+2)$ & CS & $(0, 0, \tfrac{4}{5})$ & $2.4 \times 10^{16}$ \\
$(\overline{3},1,+\tfrac{4}{3})$ & CS & $(\tfrac{1}{6}, 0, \tfrac{16}{15})$ & $2.4 \times 10^{16}$ \\
$(\overline{6},1,+\tfrac{2}{3})$ & CS & $(\tfrac{5}{6}, 0, \tfrac{8}{15})$ & $2.4 \times 10^{16}$ \\
$(1,1,0)$ & RS & $(0, 0, 0)$ & $4.1 \times 10^{16}$ \\
\hline
\end{tabular}
\caption{\label{TableSpectra}The scalar spectra corresponding to the gauge unification patterns depicted in FIGURES \ref{Figure1} and \ref{Figure2}. In the ``type'' column the type of the multiplet is encoded as follows: ``CS''=complex scalar, ``RS''=real scalar, ``GB''=Goldstone boson, ``VB''=vector boson. In the $\Delta b^{321}$ column the change of the slope of the three effective gauge couplings (i.e.,  the corresponding $\alpha^{-1}$ parameters) are given for each of the thresholds. The entries corresponding to the accidentally light scalar multiplets (in first rows) and the gauge bosons associated to the GUT-scale symmetry breaking are in boldface. The $\Delta b^{321}$ indicates the change in the slopes of the three curves in FIGUREs~\ref{Figure1} and~\ref{Figure2} upon surpassing each of the relevant energy scales.  For further details see~\cite{Bertolini:2012im}.}
\label{tab:a}
\end{table}


\end{document}